\documentclass[12pt]{article}
\usepackage{amsmath,amssymb,amsthm,latexsym,euscript,amscd, authblk}
\usepackage[english]{babel}
\usepackage{mathrsfs}
\usepackage{bm}

\newtheorem{remark}{Remark}
\newtheorem{proposition}{Proposition}
\newtheorem{corollary}{Corollary}

\title{Tomographic portrait of quantum channels}

\author[1]{G.G. Amosov\thanks{gramos@mi.ras.ru}}

\author[2]{S. Mancini\thanks{stefano.mancini@unicam.it}}

\author[3]{V.I. Man'ko\thanks{manko@lebedev.ru}}

\affil[1]{Steklov Mathematical Institute of Russian Academy of Sciences,
ul. Gubkina 8, Moscow 119991, Russia}

\affil[2]{School of Science and Technology, University of Camerino,
I-62032 Camerino, Italy,
and INFN Sezione di Perugia, I-06123 Perugia, Italy}

\affil[3]{Lebedev Physical Institute  of Russian Academy of Sciences, Leninsky pr. 53, Moscow 119991, Russia,
and Moscow Institute of Physics and Technology,
Institutski per. 9, Dolgoprudny 141700, Russia}

\begin{document}

\maketitle

\begin{abstract}
We formulate the notion of quantum channels in the framework of quantum tomography and 
address there the issue of whether such maps can be regarded as classical stochastic maps.  
In particular kernels of maps acting on probability representation of
quantum states are derived for qubit and bosonic systems.
In the latter case it results that a single mode Gaussian quantum channel corresponds to 
non-Gaussian classical channels.
\end{abstract}


\section{Introduction}

Following the general approach of \cite {MMM}, 
given a Hilbert space $H$ and a set of operators $\hat{U}({\bf x})$ acting on it, 
labelled by a $n$-dimensional real vector ${\bf x} = (x_1, x_2, . . ., x_n)$,  
we construct a complex valued function associate to an operator $\hat A$ on $H$ as
\begin{equation}\label{tomogram}
f_{\hat A}({\bf x})={\rm Tr}\left(\hat U({\bf x})\hat A\right),
\end{equation}
and called it hereafter \emph{symbol of operator} $\hat A$.
Suppose now that there exists a set of operators $\hat D({\bf x})$ on $H$ such that we can write
\begin{equation}\label{D}
\hat A=\int \hat D({\bf x})f_{\hat A}({\bf x})d{\bf x}.
\end{equation}
The requirement that the composition of maps (\ref {tomogram}) and (\ref {D}) leads to the identity operator results in
\begin{equation}\label{Completeness}
\int {\rm Tr}\left(\hat U({\bf x})\hat D({\bf y})\right)f_{\hat A}({\bf y})d{\bf y}=f_{\hat A}({\bf x}).
\end{equation}
The sets $\hat D({\bf x})$ and $\hat U({\bf x})$ are said to be quantizer and de-quantizer respectively
\footnote{Eq.\eqref{Completeness} can be regarded as the completeness relation for generalized tomographies \cite{MMT97,dAriano}.}.
If one defines the map for which the symbol of identity operator $\rm \hat I$ is equal to the
unit function, then operators $\hat U({\bf x})$ and $\hat D({\bf x})$ satisfy the conditions
\begin{equation}\label{usl}
{\rm Tr}\left(\hat U({\bf x})\right)=1,\quad \int \limits \hat D({\bf x})d{\bf x}={\rm \hat I}.
\end{equation}
 
In this framework the symbol $\omega _{\hat \rho }({\bf x})$ of a quantum state (i.e. an operator 
$\hat\rho$ on $H$ such that $\hat \rho >0$ and ${\rm Tr}\hat \rho =1$) is said to be
a quantum tomogram. We hereafter denote by $\mathfrak{T}(H)$ the set of all tomograms obtainable 
on $H$.
Taking into account (\ref {D}) we get
\begin{equation}\label{trace}
{\rm Tr}(\hat \rho )=\int {\rm Tr}\left(\hat D({\bf x})\right)f_{\hat \rho}({\bf x})d{\bf x}=1.
\end{equation}
The alternative demand to (\ref{usl}) is
\begin{equation}\label{usl2}
{\rm Tr}\left(\hat D({\bf x})\right)=1,\quad \int \limits \hat U({\bf x})d{\bf x}={\rm \hat I}.
\end{equation}
In this case the symbol of a quantum state $\hat\rho$ satisfies the relation
\begin{equation}\label {usl3}
\int \limits \omega _{\hat \rho}({\bf x})d{\bf x}=1.
\end{equation}
It should be noted that in general $\omega _{\hat \rho }({\bf x})\not\ge 0$. Hence $\omega _{\hat \rho }({\bf x})$ is not always a probability distribution. Nevertheless, it is so for important cases such as spin \cite{spintomo}, optical \cite{optomo} and symplectic \cite{symtomo} tomographies. In such contexts  
quantizer $\hat D({\bf x})$ and de-quantizer $\hat U({\bf x})$ give rise to a dual structure \cite {Amo, Dnestor}. It also should be noted that the symbol (\ref {tomogram}) becomes a
characteristic function of the quantum state $\hat \rho$ 
 whenever Weyl operators are used in place of $\hat U({\bf x})$ and $\hat D({\bf x})$ 
 \cite{Holevo}.
Moreover $f_{\hat A}({\bf x})$
can be a generalized function \cite {AmoL}.

A quantum channel $\Phi$ is a linear, completely positive trace-preserving map on the set of all states 
$\mathfrak {S}(H)$ that can be represented as \cite{K83}
\begin{equation}\label{Kraus}
\Phi (\hat\rho)=\sum \limits _i \hat {A}_i{\hat \rho}\hat {A}_i^\dag,
\quad 
\sum \limits _i \hat A_i^{\dag}\hat A_i={\rm \hat I},
\end{equation}
being $\hat {A}_i$ operators on $H$.

Any quantum channel $\Phi $ generates a map $\breve\Phi $ on the set $\mathfrak{T} (H)$
by the formula
\begin{equation}\label{chan}
\breve\Phi (\omega _{\hat \rho })({\bf x})=\omega _{\Phi (\hat \rho )}({\bf x}),
\quad 
\hat \rho \in \mathfrak{S}(H).
\end{equation}
Here we address the problem of representing (\ref{chan}) in the form
\begin{equation}\label{gen}
\breve\Phi (\omega )({\bf x})=\int {\mathcal K}({\bf x};{\bf x'})\,
\omega({\bf x'})d{\bf x'},
\quad 
\omega \in \mathfrak{T} (H),
\end{equation}
to compare quantum channels with classical stochastic maps.
The situation is considered both for finite and infinite dimensional Hilbert spaces $H$. In particular it is shown that for the bosonic Gaussian quantum channel the kernel (\ref {gen}) 
give rise to classical stochastic maps, but having a non-Gaussian form. 

By referring to \eqref{Kraus} we can write the map (\ref {gen}) with the kernel given by
\begin{equation}
 {\cal K}({\bf x} ; {\bf x'}):=\sum_i {\rm Tr}\left(\hat {U}({\bf x})\hat {A}_i\hat {D}({\bf x'})\hat  {A}_i^\dag\right).
\end{equation}

If (\ref {usl2}) is satisfied and
\begin{equation}\label{rav}
 \int  \,  {\cal K}({\bf x} ; {\bf x}') \, {\rm d}{\bf x}' = 1,
\end{equation}
then the map defined by (\ref {gen}) has the property
$$
\int \breve \Phi (\omega )({\bf x})d{\bf x}=\int \omega ({\bf x})d{\bf x}
$$
which is equivalent to preserving the trace for $\Phi$. Nevertheless (\ref {rav}) is not take
place in general because the set $\mathfrak {T}(H)$ can not coincide with the set of all probability
distributions \cite {AmoL}. Moreover, 
${\cal K}({\bf x};{\bf x}') \not\ge 0$.
Thus $ {\cal K}({\bf x} ; {\bf x}')$ is not in general a conditional probability.  
Analogously the unitality of a channel $\Phi $, i.e. $\Phi(\frac {1}{dimH}\hat{I})=\frac {1}{dimH}\hat{I}$, 
is not equivalent to claim
\begin{equation}
\int\,  {\cal K}({\bf x} ; {\bf x}')\,  {\rm d}{\bf x} =1.
\end{equation}
Taking into account that $\Phi $ is completely positive iff
\begin{equation}
\sum \limits _{j,k} \langle \xi _j |\Phi \left(|\eta _j\rangle\langle \eta _k|\right)|\xi _k\rangle \ge 0,
\quad
\forall |\xi _j\rangle,|\eta _k\rangle\in H,
\end{equation}
we obtain the necessary and sufficient condition on ${\cal K}$ to determine
a quantum channel in tomographic representation. That is
\begin{equation}\label{cond}
\sum \limits _{j,k}\int \int {\cal K}({\bf x};{\bf x'})
\langle \xi _j|\hat D({\bf x})|\xi _k\rangle\langle \eta _k|\hat U({\bf x'})|\eta _j\rangle
d{\bf x}d{\bf x'}\ge 0,
\quad
 \forall |\xi _j\rangle,|\eta _k\rangle\in H.
\end{equation}

\section{Qubit channels}

The qubit (spin-$\frac{1}{2}$) tomogram is given by \cite{spintomo, Fil} 
\begin{equation}
\label{eq:FS}
w_{\hat \rho}({\bf x})=w({\bf x})={\rm Tr}\left(\hat{\rho}\,\hat{U}({\bf x})\right),
\end{equation}
where ${\bf x}:=(m,\alpha,\beta)$. Here $m=\pm\frac{1}{2}$ are the two possible outcomes of the spin measurement performed along the direction $(\sin\alpha\cos\beta,\sin\alpha\sin\beta,\cos\beta)$ determined by
the Euler angles $\alpha,\beta$.

The operators $\hat{U}({\bf x})$ read
\begin{equation}\label{u}
\hat{U}({\bf x})=\frac{1}{2}
\begin{pmatrix}
1 & 0 \\
0 & 1
\end{pmatrix}
+m \begin{pmatrix}
\cos\beta & -e^{i\alpha}\sin\beta \\
-e^{-i\alpha}\sin\beta & -\cos\beta
\end{pmatrix}.
\end{equation}

The tomograms satisfy the normalization conditions
\begin{equation}
\sum_{m=-1/2}^{1/2} w(m,\alpha,\beta)=1,
\qquad
\frac{1}{2\pi}\int_0^{2\pi} \int_0^\pi \, 
w(m,\alpha,\beta)\,
\sin\beta \,{\rm d} \beta\, 
{\rm d}\alpha=1.
\end{equation}

Equation \eqref{eq:FS} can be inverted by expressing the density operator in terms of tomograms as 
\begin{equation}
\hat{\rho}=\int  \, \hat{D}({\bf x}) \, w({\bf x}) {\rm d} {\bf x},
\end{equation}
where 
\begin{equation}
\int {\rm d}{\bf x}:=\sum_{m=-1/2}^{1/2}\frac{1}{2\pi}\int_0^{2\pi} {\rm d}\alpha\int_0^\pi \sin\beta \,{\rm d} \beta ,
\end{equation}
and
\begin{equation}
\hat{D}({\bf x}):=3 \hat{U}({\bf x})-{\rm \hat I}.
\end{equation}

A channel $\Phi :\frak {S}(\mathbb{C}^2)\to \frak {S}(\mathbb{C}^2)§$ defines the linear map $\breve\Phi $ on the set $\mathfrak{T}(\mathbb{C}^2)$ of spin-$\frac{1}{2}$ tomograms by the formula 
\begin{equation}\label{map}
\breve\Phi (w_{\hat \rho})(m,\alpha ,\beta )=w_{\Phi (\hat \rho)}(m,\alpha,\beta ).
\end{equation}
The matrix (\ref {u}) can be represented as follows
\begin{equation}\label{u2}
\hat U({\bf x})=\frac {1}{2}{\rm \hat I}-m\cos\alpha\sin\beta \,\hat\sigma _x
-m\sin\alpha\sin\beta \,\hat \sigma_y+m\cos\beta \, \hat\sigma _z,
\end{equation}
where $\hat\sigma _x$, $\hat\sigma _y$, $\hat\sigma _z$ are the standard Pauli operators.
 Thus, to determine $\breve\Phi $ one should
check the action of a conjugate map $\Phi ^*$, that is ${\rm Tr}\left(\hat\rho \Phi ^*(\hat\sigma )\right)
={\rm Tr}\left(\Phi (\hat\rho )\hat\sigma\right)$, on (\ref {u2}).

\subsection{Unital qubit channel}

All unital qubit channels $\Phi :\mathfrak {S}(\mathbb{C}^2)\to \mathfrak {S}(\mathbb{C}^2)$ 
are mixture of unitary channels, i.e. there are unitary operators 
$\hat{U}_j :\mathbb{C}^2\to\mathbb{C}^2$
such that
\begin{equation}\label{uch}
\Phi (\hat \rho )=\sum \limits _j\pi _j\hat{U}_j\hat \rho \hat{U}_j^*,
\end{equation}
$\pi _j\ge 0,\ \sum \limits _j\pi _j=1$. Moreover, picking up unitaries $\hat U,\hat V:\mathbb{C}^2\to\mathbb{C}^2$ we can obtain the representation (\ref {uch}) for the channel
$\Psi (\hat \rho )=\hat U\Phi (\hat V\hat \rho \hat V^*)\hat U^*$ with $\hat U_j\in SU(2)$.

Let us write ${\bf x}=(m,\overline n)$, where $\overline n:=(\cos\alpha \sin\beta,\sin\alpha \sin\beta, \cos\beta )$.
It follows from (\ref {u2}) and (\ref {uch}) that
\begin{equation}\label{kraus2}
\breve\Phi (w)(m,\overline n)=\sum \limits _j\pi _jw(m,\hat V_j\overline n),
\end{equation}
where $\hat V_j\in O(3)$.

Given a unital qubit channel $\Phi :\mathfrak{S}(\mathbb{C}^2)\to\mathfrak{S}(\mathbb{C}^2)$, there exist unitary operators $\hat U,\hat V:\mathbb{C}^2\to\mathbb{C}^2$ 
such that
\begin{equation}\label{unital}
\Psi (\hat \rho)=\hat U\Phi (\hat V\hat \rho \hat V^*)\hat U^*
=\pi _0\hat \rho +\pi _x\hat\sigma _x\hat\rho \hat\sigma _x+\pi _y\hat\sigma _y\hat\rho \hat\sigma _y
+\pi _z\hat\sigma _z\hat\rho \hat\sigma _z,\quad \hat\rho \in \mathfrak {S}(\mathbb{C}^2),
\end{equation}
where $\{\pi _0,\pi _x,\pi _y,\pi _z\}$ is a probability distribution. Thus, it suffices to study only channels
$\Psi $ of the form (\ref {unital}). 
Denote by $\breve \Sigma _a$ the unitary quantum channel implemented by the Pauli matrix $\hat \sigma _a$, i.e.
$$
\breve \Sigma _a(\hat \rho )=\hat \sigma _a\hat \rho \hat \sigma _a,
\quad \hat \rho \in \mathfrak {S}(\mathbb{C}^2), 
$$
with $a\in\{x,y,z\}$.

\begin{proposition}{Proposition} The linear maps $\breve\Sigma _x$, $\breve\Sigma _y$ and $\breve\Sigma _z$ act on the set $\mathfrak{T}(\mathbb{C}^2)$ of qubit tomograms as follows
\begin{eqnarray}\label{maps}
\breve\Sigma _x &:& w(m,\alpha ,\beta )\to w\left(m,\alpha -\frac {\pi }{2},\beta +\frac {\pi }{2}\right)
\notag\\
\breve\Sigma _y &:& w(m,\alpha ,\beta )\to w\left(m,\alpha +\frac {\pi }{2},\beta +\frac {\pi }{2}\right)
\notag\\
\breve\Sigma _z &:& w(m,\alpha ,\beta )\to w\left(m,\alpha ,\beta -\frac {\pi }{2}\right).
\end{eqnarray}
\end{proposition}

Proof. It is 
$$
\breve \Sigma _a(w_{\hat \rho})({\bf x})={\rm Tr}(\hat \sigma _a\hat \rho \hat \sigma _a\hat U({\bf x}))={\rm Tr}(\hat \rho \hat \sigma _a\hat U({\bf x})\hat \sigma _a),
$$
$a\in \{x,y,z\}$.
Taking into account (\ref {u2}) we get
$$
\hat \sigma _x\hat U({m,\alpha ,\beta  })\hat \sigma _x=\hat U\left (m,\alpha -\frac {\pi }{2},\beta +\frac {\pi }{2}\right),
$$
$$
\hat \sigma _y\hat U(m,\alpha ,\beta )\hat \sigma _y=\hat U\left (m,\alpha +\frac {\pi }{2},\beta +\frac {\pi }{2}\right),
$$
$$
\hat \sigma _z\hat U(m,\alpha ,\beta )\hat \sigma _z=\hat U\left (m,\alpha ,\beta -\frac {\pi }{2} \right).
$$

\hfill$\blacksquare$

\begin {corollary}{Corollary} The linear map $\breve\Phi $ on the set $\mathfrak{T}(\mathbb{C}^2)$ of qubit tomograms is associated with a unital quantum channel iff
it is (up to unitary equivalence) a convex linear combination of the identity map and the three maps (\ref {maps}).
\end{corollary}

Proof. It immediately follows from the representation of unital channel in the form (\ref {unital}).

\hfill$\blacksquare$

\begin{proposition} {Proposition} The maps $\breve\Sigma _x$, $\breve\Sigma _y$ and $\breve\Sigma _z$ can be represented in the form of integral
operators
$$
\breve \Sigma _a (w)({\bf x})=\int K_{a}({\bf x};{\bf x'})w({\bf x'})\,d{\bf x'},\quad a\in \{x,y,z\},
$$
with the kernels defined by the formula
$$
K_{x}({\bf x};{\bf x'})=\frac {1}{2}\delta _{mm'}(1+3\cos \alpha \sin \beta \cos \alpha '\sin\beta '-3\sin \alpha \sin \beta \sin \alpha '\sin \beta '-3\cos \beta \cos \beta '),
$$
$$
K_{y}({\bf x};{\bf x'})=\frac {1}{2}\delta _{mm'}(1-3\cos \alpha \sin \beta \cos \alpha '\sin\beta '+3\sin \alpha \sin \beta \sin \alpha '\sin \beta '-3\cos \beta \cos \beta '),
$$
$$
K_{z}({\bf x};{\bf x'})=\frac {1}{2}\delta _{mm'}(1-3\cos \alpha \sin \beta \cos \alpha '\sin\beta '-3\sin \alpha \sin \beta \sin \alpha '\sin \beta '+3\cos \beta \cos \beta ').
$$
\end{proposition}

Proof. Let us define the inner product by the formula
\begin{equation}\label{inn}
(f,g):= \int \limits _0^{2\pi } \int \limits _0^{\pi } \overline {f(\alpha ,\beta )}g(\alpha ,\beta )
\, \sin \beta d\beta\,
d\alpha.
\end{equation}
Then, the functions

\begin{equation}\label{functions}
f_0(\alpha ,\beta )=1,\quad 
f_1(\alpha ,\beta )=\cos \alpha \sin\beta ,\quad 
f_2(\alpha ,\beta )=\sin\alpha \sin\beta ,\quad
f_3(\alpha ,\beta )=\cos \beta,
\end{equation}
become orthogonal with respect to (\ref {inn}). Moreover,
$$
||f_0||^2=2,\ ||f_1||^2=||f_2||^2=||f_3||^2=\frac {2}{3}.
$$
To fullfil the transformation from Proposition 1 one can construct the kernels using this set of orthogonal functions.

\hfill$\blacksquare$

\begin{remark} The kernels determined in Proposition 3 are not positive definite. Thus, the maps $\breve \Sigma _x,\breve \Sigma _y$ and $\breve \Sigma _z$ are
not classical channels. 
\end{remark}

\subsection{Non-unital qubit channels}

Given a qubit channel $\Phi:\mathfrak {S}(\mathbb{C}^2)\to \mathfrak {S}(\mathbb{C}^2)$ there exist unitaries 
$\hat U,\hat V:\mathbb{C}^2\to\mathbb{C}^2$, and a set of real numbers $(t_x,t_y,t_z,\lambda _x,\lambda _y,\lambda _z)$ 
such that
\begin{equation}\label {general}
\Psi (\hat \rho)=\hat U\Phi (\hat V\hat \rho \hat V^*)\hat U^*=\frac {1}{2}\left( {\rm \hat I}+(t_x+\lambda _xa_x)\hat \sigma _x+(t_y+\lambda _ya_y)\hat \sigma _y+(t_z+\lambda _za_z)\hat \sigma _z\right),
\end{equation}
where
$$
\hat \rho =\frac {1}{2}({\rm \hat I}+a_x\hat \sigma _x+a_y\hat \sigma_y+a_z\hat \sigma _z).
$$
The image of the Bloch sphere of pure states under a map of the form (\ref {general}) is the ellipsoid
$$
\left (\frac {x_1-t_1}{\lambda _1}\right )^2+\left (\frac {x_2-t_2}{\lambda _2}\right )^2+\left (\frac {x_3-t_3}{\lambda _3}\right )^2=1.
$$
The conditions on the parameters $(t_x,t_y,t_z,\lambda _x,\lambda _y,\lambda _z)$ for which $\Psi $ is a channel are quite complicated and derived in \cite {ruskai}.

The extreme points of the set (\ref {general}) for non-unital case correspond (up to unitary equivalence) to 
\begin{equation}\label{extreme}
t_x=t_y=0,\quad \lambda _z=\lambda _x\lambda _y,\quad t_z^2=(1-\lambda _x^2)(1-\lambda _y^2).
\end{equation}

For the conjugate map we obtain
\begin{equation}\label {general2}
\Psi ^*(\hat \rho)=\hat U\Phi (\hat V\hat \rho \hat V^*)\hat U^*=\frac {1}{2}\left((1+t_xa_x+t_ya_y+t_za_z){\rm \hat I}+\lambda _xa_x\hat \sigma _x+\lambda _ya_y\hat \sigma _y+\lambda _za_z\hat \sigma _z\right).
\end{equation}
Substituting (\ref {u2}) into (\ref {general2}) we get
\begin{eqnarray}\label{u3}
\Psi ^*(\hat U({\bf x}))&=&\frac {1}{2}\left(1-t_xm\cos\alpha\sin\beta -t_ym\sin\alpha\sin\beta +t_zm\cos\beta \right){\rm \hat I} \notag\\
&-&\lambda_xm\cos\alpha\sin\beta \,\hat\sigma _x
-\lambda_ym\sin\alpha\sin\beta \,\hat \sigma_y+\lambda_zm\cos\beta \,\hat\sigma _z.
\end{eqnarray}

\begin{proposition}{Proposition} The map (\ref {map}) associated with the channel (\ref {general}) can be represented in the form of integral operator
$$
\breve \Psi (w)({\bf x})=\int {\cal K}({\bf x};{\bf x'})w({\bf x'})\,d{\bf x'},
$$
with the kernel
\begin{eqnarray}\label{K33}
{\mathcal K}({\bf x};{\bf x'})&=&\frac {\delta _{mm'}}{2}(1-m\cos \alpha \sin\beta t_x-m\sin\alpha \sin\beta t_y+m\cos \alpha t_z)\notag\\
&+&\frac {3}{2}\delta _{mm'}\left(-\cos \alpha \sin\beta \cos\alpha '\sin\beta ' \lambda _x-\sin\alpha \sin\beta \sin\alpha '\sin \beta ' \lambda _y+\cos \beta \cos \beta '\lambda _z\right).\notag\\ 
\end{eqnarray}
\end{proposition}

Proof. Following the idea of proof in Proposition 3, take into account that the functions (\ref {functions}) are orthogonal. Then, by means of them we construct the kernel
corresponding to the transformation ({\ref {u3}).

\hfill$\blacksquare$

\begin{remark} Like for unital channels the kernel \eqref{K33} is not positive definite and the map $\hat \Psi $ is not a classical channel determined by conditional probabilities. 
\end{remark}

\section{One-mode Bosonic channel}

 In this section we shall move to the framework of optical homodyne tomography of a single-mode radiation field (see e.g. Refs.\cite {rev1,rev2}).
The optical tomogram $\omega _{\hat \rho }(x,\varphi )$ of a state $\hat\rho $ in $L^2(\mathbb{R})$ is given by the formula \cite{optomo}
\begin{equation}\label{opttom}
\omega (x,\varphi)=\omega _{\hat \rho}(x,\varphi )={\rm Tr}\left(\hat \rho\, \delta\left(x-\cos \varphi\hat Q-\sin\varphi \hat P\right)\right),
\end{equation}
where $\hat Q,\hat P$ are the canonical conjugate quadratures operators and $x\in\mathbb{R},\varphi\in[0,2\pi]$.
The characteristic function $F(q,p)$ relative to $\hat \rho $ is defined as  
\begin{equation}\label{charact}
F(q,p)=F_{\hat \rho}(q,p)={\rm Tr}\left(\hat \rho \, e^{i(q \hat Q+p \hat P)}\right).
\end{equation}

The optical tomogram $\omega (x,\varphi )$ is connected with the characteristic function $F(q,p)$ as follows
\begin{equation}\label {first}
F(t\cos \varphi ,t\sin\varphi)=\int \limits _{\mathbb R}e^{itx}\omega (x,\varphi)dx,
\end{equation}
\begin{equation}\label{second}
\omega (x,\varphi )=\frac {1}{2\pi }\int \limits _{\mathbb R}e^{-ixt}F(t\cos\varphi,t\sin\varphi)dt.
\end{equation}

Following up \eqref{chan}, consider a map $\breve\Phi $ on the set of optical tomograms given} by the formula
\begin{equation}\label{mapa}
\breve\Phi (\omega _{\hat \rho})(x,\varphi )=\omega _{\Phi (\hat \rho)}(x,\varphi).
\end{equation}

Below we shall deal with quantum Gaussian channels, widely used in quantum information (see e.g. \cite{rev3}).

\subsection{Covariant channel}

Let us take a one-mode covariant Bosonic channel $\Phi $ transforming the characteristic function  $F(q,p)$
by the formula \cite {Holevo}
\begin{equation}\label{boson}
F(q,p)\to F(kq,kp)e^{-\frac {\alpha (q^2+p^2)}{2}},
\end{equation}
being
$$
k\ge 0,\quad k\neq 1,\quad \alpha \ge \frac {|k^2-1|}{2}.
$$

\begin{proposition} {Proposition} The map (\ref {mapa}) associated with the Bosonic channel (\ref {boson}) can be represented as
an integral operator with a Gaussian kernel
\begin{equation}\label{GKcov}
\breve\Phi (\omega )(x,\varphi )=\frac {1}{\sqrt {2\pi \alpha }}\int \limits _{\mathbb R}e^{-\frac {(x-kx')^2}{2\alpha }}\omega (x',\varphi )dx'.
\end{equation}
\end{proposition}

Proof.
Taking into account the relations (\ref {first}), (\ref {second}) and (\ref {boson}) we get
$$ 
\breve\Phi (\omega )(x,\varphi )=\frac {1}{2\pi }\int \limits _{\mathbb R}e^{-ixy}e^{-\frac {\alpha y^2}{2}}\int \limits _{\mathbb R}e^{ikyx'}\omega (x',\varphi )dx'dy.
$$ 
Changing the order of integration we arrive at
$$
\frac {1}{2\pi}\int \limits _{\mathbb R}e^{i(kx'-x)y}e^{-\frac {\alpha y^2}{2}}dy=\frac {1}{\sqrt {2\pi \alpha}}e^{-\frac {(kx'-x)^2}{2\alpha }}.
$$
\hfill$\blacksquare$

\begin{remark} The kernel ${\cal K}(x,\varphi;x',\varphi')=\frac {1}{\sqrt {2\pi \alpha}}
e^{-\frac {(x-kx')^2}{2\alpha }}\delta(\varphi'-\varphi)$ resulting from \eqref{GKcov} is positive definite and $\int {\cal K}(x,\varphi;x',\varphi')dxd\varphi=1$, $\int {\cal K}(x,\varphi;x',\varphi')dx'd\varphi'=\frac {1}{k}$. 
Hence the map \eqref{GKcov} results stochastic, but not bi-stochastic. As matter of fact ${\cal K}(x,\varphi;x',\varphi')$ does not represent a conditional probability distribution. 
\end{remark}

\subsection{Contravariant channel}

Let us now take a one-mode contravariant Bosonic channel $\Phi $ transforming the characteristic function  $F(q,p)$ by the formula \cite {Holevo}
\begin{equation}\label{boson2}
F(q,p)\to F(kq,-kp)e^{-\frac {\alpha (q^2+p^2)}{2}},
\end{equation}
being
$$
k\ge 0,\quad \alpha \ge \frac {k^2+1}{2}.
$$

\begin{proposition} {Proposition}  The map (\ref {mapa}) associated with the Bosonic channel (\ref {boson2}) can be represented as
an integral operator with a Gaussian kernel 
\begin{equation}\label{GKcont}
\breve\Phi (\omega )(x,\varphi )=\frac {1}{\sqrt {2\pi \alpha}}\int \limits _{\mathbb R}e^{-\frac {(x-kx')^2}{2\alpha }}\omega \left (x',\varphi -\frac {\pi}{2} \right )dx'.
\end{equation}
\end{proposition}

Proof.
Taking into account the relations (\ref {first}), (\ref {second}) and (\ref {boson2}) we get
$$
\breve\Phi (\omega )(x,\varphi )=\frac {1}{2\pi }\int \limits _{\mathbb R}e^{-ixy}e^{-\frac {\alpha y^2}{2}}\int \limits _{\mathbb R}e^{ikyx'}\omega 
\left (x',\varphi -\frac{\pi }{2}\right )dx'dy.
$$ 
Changing the order of integration we arrive at
$$
\frac {1}{2\pi}\int \limits _{\mathbb R}e^{i(kx'-x)y}e^{-\frac {\alpha y^2}{2}}dy=\frac {1}{\sqrt {2\pi \alpha }}e^{-\frac {(kx'-x)^2}{2\alpha }}.
$$
\hfill$\blacksquare$

\begin{remark} The kernel 
${\cal K}(x,\varphi;x',\varphi')=\frac {1}{\sqrt {2\pi \alpha }}
e^{-\frac {(x-kx')^2}{2\alpha }}\delta(\varphi'-\varphi+\pi/2)$ resulting
from \eqref{GKcont} is positive definite and 
$$\int {\cal K}(x,\varphi;x',\varphi')dxd\varphi=1,$$ 
$$\int {\cal K}(x,\varphi;x',\varphi')dx'd\varphi'=\frac {1}{k}.$$ 
Hence the map \eqref{GKcont} results stochastic, but not bi-stochastic. As matter of fact ${\cal K}(x,\varphi;x',\varphi')$ does not represent a conditional probability distribution.
\end{remark}

\subsection{The representation on the plane}

Following \cite{Amo} let us define the function $\Omega (x,y)$ on the plane ${\mathbb R}^2$ in polar
coordinates by the formula 
\begin{equation}\label{O}
{\Omega }(r\cos\varphi,r\sin\varphi)={\Omega } _{\hat \rho }(r\cos\varphi,r\sin\varphi):=\frac {1}{r}\omega (r,\varphi).
\end{equation}
Then,
$$
{\Omega }(x,y)\ge 0,\quad \frac {1}{2\pi}\int \limits_{{\mathbb R}^2} {\Omega }(x,y)dxdy=1,
$$
hence ${\Omega }$ results a probability distribution function on ${\mathbb R}^2$.
It follows from the definition (\ref {O}) that the characteristic function can be reconstruct from
(\ref {O}) by the formula
\begin{equation}\label{F}
F(t\cos\varphi ,t\sin\varphi )=\int \limits _0^{+\infty}re^{itr}\Omega (r\cos\varphi ,r\sin\varphi )dr.
\end{equation}
Consider now the linear map on the set of functions (\ref {O}) 
\begin{equation}\label{plain}
\breve\Phi ({\Omega} _{\hat \rho})(x,y)={\Omega } _{\Phi (\hat \rho )}(x,y). 
\end{equation}

\begin{proposition}{Proposition} The map (\ref {plain}) associated with the Bosonic channel (\ref {boson}) is the integral operator
$$
\breve\Phi ({\Omega })(x,y)=\int \limits _{{\mathbb R}^2}{\cal K}(x,y;x',y'){\Omega }(x',y')dx'dy',
$$
with the kernel
\begin{equation}\label {form}
{\cal K}(x,y;x',y')=\frac {1}{\sqrt {2\pi \alpha}}\exp\left (-\frac {(x-kx')^2+(y-ky')^2}{2\alpha }\right )\delta _{x,y}(x',y'),
\end{equation}
where
$$
\langle \delta _{x,y}, \psi \rangle :=\frac {1}{\sqrt {x^2+y^2}}\int \limits _{0}^{+\infty}r\psi \left (r\frac {x}{\sqrt {x^2+y^2}},r\frac {y}{\sqrt {x^2+y^2}}\right )dr.
$$
\end{proposition}

Proof. It is
$$
\breve \Phi (\Omega )(\rho \cos\varphi ,\rho \sin\varphi)=\frac {1}{2\pi\rho }\int \limits _{\mathbb R}e^{-it\rho}e^{-\alpha \frac {t^2}{2}} \int 
\limits _0^{+\infty }re^{iktr}\Omega (r\cos \varphi ,r\sin\varphi)drdt.
$$
Changing the order of integration we get
$$
\frac {1}{2\pi}\int \limits _{\mathbb R}e^{it(kr-\rho )}e^{-\alpha \frac {t^2}{2}}dt=\frac {1}{\sqrt {2\pi \alpha }}e^{-\frac {(\rho -kr)^2}{2\alpha }},
$$
and
$$
\breve \Phi (\Omega )(\rho \cos\varphi ,\rho \sin\varphi)=\frac {1}{\sqrt {2\pi \alpha }\rho}\int \limits _{0}^{+\infty }
re^{-\frac {(\rho -kr)^2}{2\alpha}}\Omega (r\cos \varphi ,r\sin\varphi)dr.
$$
Substituting $x=\rho \cos\varphi,\ y=\rho \sin\varphi $ we obtain
$$
\breve \Phi (\Omega )(x,y)=\frac {1}{\sqrt {2\pi \alpha (x^2+y^2)}}\int \limits _{0}^{+\infty }re^{-\frac {(\sqrt {x^2+y^2}-kr)^2}{2\alpha }}\Omega \left (r\frac {x}{\sqrt {x^2+y^2}},
r\frac {y}{\sqrt {x^2+y^2}}\right )dr.
$$

\hfill$\blacksquare$

\begin{remark} It is worth remarking that the same conclusion of 
Proposition 7 can be drawn for contravariant channels 
simply changing $(x,y)$ to $(y,-x)$ for $\Omega$.
\end{remark}

\begin{remark} The kernel (\ref {form}) is positive definite and $\int {\cal K}(x,y;x',y')dxdy=1$ and
$\int {\cal K}(x,y;x',y')dx'dy'=\frac{1}{k}$. 
Hence the map \eqref{plain} results stochastic, but not bi-stochastic. As matter of fact ${\cal K}(x,\varphi;x',\varphi')$ does not represent a conditional probability distribution.
Anyway, the one-mode bosonic channel (be either covariant or contravariant) can be intended through the representation on the plane as a two-mode classical channel, i.e. acting on probability distribution functions on $\mathbb{R}\times\mathbb{R}$.
This is in contrast to the map \eqref{mapa} where the argument is defined on $\mathbb{R}\times [0,2\pi]$.
\end{remark}

\section{Conclusion}

In conclusion, we have formulate the notion of quantum channel in the framework of quantum tomography, that is as a map acting on probability representation of
quantum states (tomograms). 
Kernels for such maps were derived for qubit and bosonic systems.
They show the existence of cases in which a quantum channel can be regarded as a classical stochastic map. In particular this happens for the one-mode bosonic channel  that corresponds to 
classical channels, though non-Gaussian. 

The present study paves the way for finding further correspondences between quantum channels and classical stochastic maps.
This could be helpful for characterizing the information transmission capabilities of quantum channels without the necessity of resorting to regularization procedures \cite{reg}.
In fact it is known that (unlike quantum channels) classical channels admit single letter formula for capacity \cite{SW49}.

\end{document}